\documentclass[twocolumn,showpacs,preprintnumbers]{revtex4}
\usepackage{graphicx}% Include figure files
\usepackage{dcolumn}% Align table columns on decimal point
\usepackage{bm}% bold math
\usepackage{amssymb}
\usepackage{amsmath}
\usepackage{epsfig}    
\usepackage{here}
\def\beq{\begin{equation}}
\def\eeq{\end{equation}}
\def\bea{\begin{array}}
\def\eea{\end{array}}
\def\be{\begin{equation}}
\def\ee{\end{equation}}
\def\ba{\begin{eqnarray}}
\def\ea{\end{eqnarray}}

\def\to{\rightarrow}

\def\[{\left[}
\def\]{\right]}
\def\({\left(}
\def\){\right)}

\def\sm0{{\widetilde{m}_0}}

\def\U1em{{U(1)_{\rm em}}}
\def\to{\rightarrow}

\def\sq2{\sqrt{2}}

\def\ee{e^+e^-}

\def\End{\end{document}}
\def\fsl#1{\setbox0=\hbox{$#1$}                 % set a box for #1 
   \dimen0=\wd0                                 % and get its size
   \setbox1=\hbox{/} \dimen1=\wd1               % get size of /
   \ifdim\dimen0>\dimen1                        % #1 is bigger
      \rlap{\hbox to \dimen0{\hfil/\hfil}}      % so center / in box
      #1                                        % and print #1
   \else                                        % / is bigger
      \rlap{\hbox to \dimen1{\hfil$#1$\hfil}}   % so center #1
      /                                         % and print /
   \fi}

\begin{document}                                                              
%\draft
%@twocolumnfalse\endcsname

\title{Doubly-charged scalar bosons from the doublet}%
\author{%
{\sc Mayumi Aoki\,$^1$, Shinya Kanemura\,$^2$,
   and Kei Yagyu\,$^2$}
}
\affiliation{%
%\address{\vspace*{5mm}
\vspace*{2mm} 
$^1$Institute~for~Theoretical~Physics,~Kanazawa~University, 
Kanazawa~920-1192, ~Japan\\
$^2$Department of Physics, University of Toyama, 3190 Gofuku, 
Toyama 930-8555, Japan\\ 
}
%\maketitle

%\vspace*{5mm} 
\begin{abstract}
%\hspace*{-0.35cm}
We consider the extended Higgs models, in which one of the isospin 
doublet scalar fields carries the hypercharge $Y=3/2$. 
Such a doublet field $\Phi_{3/2}$ is composed of a doubly charged scalar 
boson as well as a singly charged one.  
We first discuss a simple model with $\Phi_{3/2}$ (Model~I),  
and study its collider phenomenology at the LHC.  
We then consider a new model for radiatively generating neutrino 
masses with a dark matter candidate (Model~II), 
in which $\Phi_{3/2}$ and an extra $Y=1/2$ doublet 
as well as vector-like singlet fermions carry 
the odd quantum number for an unbroken discrete $Z_2$ symmetry. 
We also discuss the neutrino mass model (Model~III),  
in which  the exact $Z_2$ parity  in Model~II is softly broken.
It is found that the doubly charged scalar bosons in these models 
show different phenomenological aspects from 
those which appear in models with a $Y=2$ isospin 
singlet field or a $Y=1$ triplet one.  
They could be clearly distinguished at the LHC.
\pacs{\, 12.60.Fr, 14.80.Fd   \hfill~~[\today] }
\end{abstract}

\maketitle

\setcounter{footnote}{0}
\renewcommand{\thefootnote}{\arabic{footnote}}

\section{Introduction} 

Physics of electroweak symmetry breaking remains 
the last unknown part of the standard model (SM), 
and currently Higgs boson searches are underway 
at the Tevatron and the LHC. 
On the other hand, some phenomena beyond the SM such as neutrino 
oscillation and the existence of dark matter have been confirmed 
by experiments, and various 
models beyond the SM 
have been proposed to explain these phenomena. 
In such new physics models, non-minimal Higgs sectors are 
often introduced. 

Charged scalar states generally appear in extended Higgs sectors. 
Although such charged states are in themselves new physics phenomena,  
by measuring their property the Higgs sector and thus 
the direction of new physics could be determined.
In particular, doubly charged scalar states are a clear signature 
for Higgs sectors with 
non-standard representations such as isospin 
triplet fields with the hypercharge $Y=1$ and
singlet fields with $Y=2$. 
Their phenomenological aspects strongly depend on the 
model, so that they can give important information to distinguish 
these models. 
The triplet fields are introduced in various models 
such as the left-right model~\cite{LRmodel}, 
the littlest Higgs model~\cite{LHmodel} and 
the Type II seesaw model~\cite{TypeIIseesaw}, 
while the singlet fields appear in various models 
for the grand unification and also in radiative seesaw 
models~\cite{ZBmodel}. 
Phenomenology of these doubly charged states thus has  
been studied extensively. 

There are, however, other representations which contain doubly 
charged scalar states. The simplest example 
is the isospin doublet scalar field $\Phi_{3/2}$ with $Y=3/2$.  
Its phenomenology has hardly been studied. 
The field $\Phi_{3/2}$ may appear in a model 
for the coupling unification~\cite{gunion}, 
or, as we shall discuss later in details, it 
can be used to build new versions of simple 
radiative seesaw models with  
a dark matter candidate~\cite{knt,ma,aks} or 
without it~\cite{zee,ZBmodel}. 

In this Letter, we discuss phenomenology of 
$\Phi_{3/2}$ in renormalizable theories.   
Contrary to the $Y=1$ triplet field $\Delta$ 
as well as the $Y=2$ singlet $S^{++}$, 
the Yukawa coupling between $\Phi_{3/2}$ 
and charged leptons is protected by the chirality.  
In addition, the component 
fields of $\Phi_{3/2}$ are both charged and do not receive a 
vacuum expectation value (VEV) as long as electric charge is conserved.  
Hence, the field decays via the mixing with the other scalar 
representations which can decay into the SM particles 
or via some higher order couplings.
This characteristic feature of $\Phi_{3/2}$ 
would give discriminative predictions at collider experiments. 
We therefore first study collider signatures of 
$\Phi_{3/2}$ at the LHC in the model (Model~I) of 
an extension from the SM with an extra $Y=1/2$ 
doublet and $\Phi_{3/2}$.  
We then present a new TeV scale model with  
$\Phi_{3/2}$, an extra $Y=1/2$ doublet and  
vector-like singlet fermions (Model~II), 
which contains a mechanism for radiatively 
generating tiny neutrino masses with a dark matter candidate 
under the exact $Z_2$ parity.  
We then briefly discuss phenomenology of the model under 
the current data from experiments 
for neutrino oscillation, lepton flavor violation (LFV), 
and dark matter. 
We also discuss the neutrino mass model (Model~III),  
in which the $Z_2$ parity  in Model~II is softly broken.

\section{Model I} 
%\underline{\bf Model I} 
The simplest model, 
where $\Phi_{3/2}$ is just added to the SM,  
can decay 
into SM particles only if lepton-number violating  
higher order operators are introduced~\cite{gunion2}. 
Thus, we here consider the model in which 
$\Phi_{3/2}$ is added to the model with two $Y=1/2$ Higgs 
doublet fields $\phi_1$ and $\phi_2$ (Model~I). 
The singly charged scalar state in $\Phi_{3/2}$
can decay into the SM particles via the mixing with 
the physical charged state from the $Y=1/2$ doublets. 
This model can be regarded as an effective theory of Model~III 
which we discuss later, or 
it may be that of the model with an additional heavier $\Delta$,    
in which the gauge coupling unification would be possible. 
In order to avoid flavor changing neutral current, 
a softly-broken $Z_2$ symmetry is imposed~\cite{gw}, under which 
the scalar fields are transformed as $\phi_1\to \phi_1$, 
$\phi_2 \to - \phi_2$, and $\Phi_{3/2} \to - \Phi_{3/2}$. 

\begin{figure}[t]
\begin{center}
\includegraphics[width=60mm]{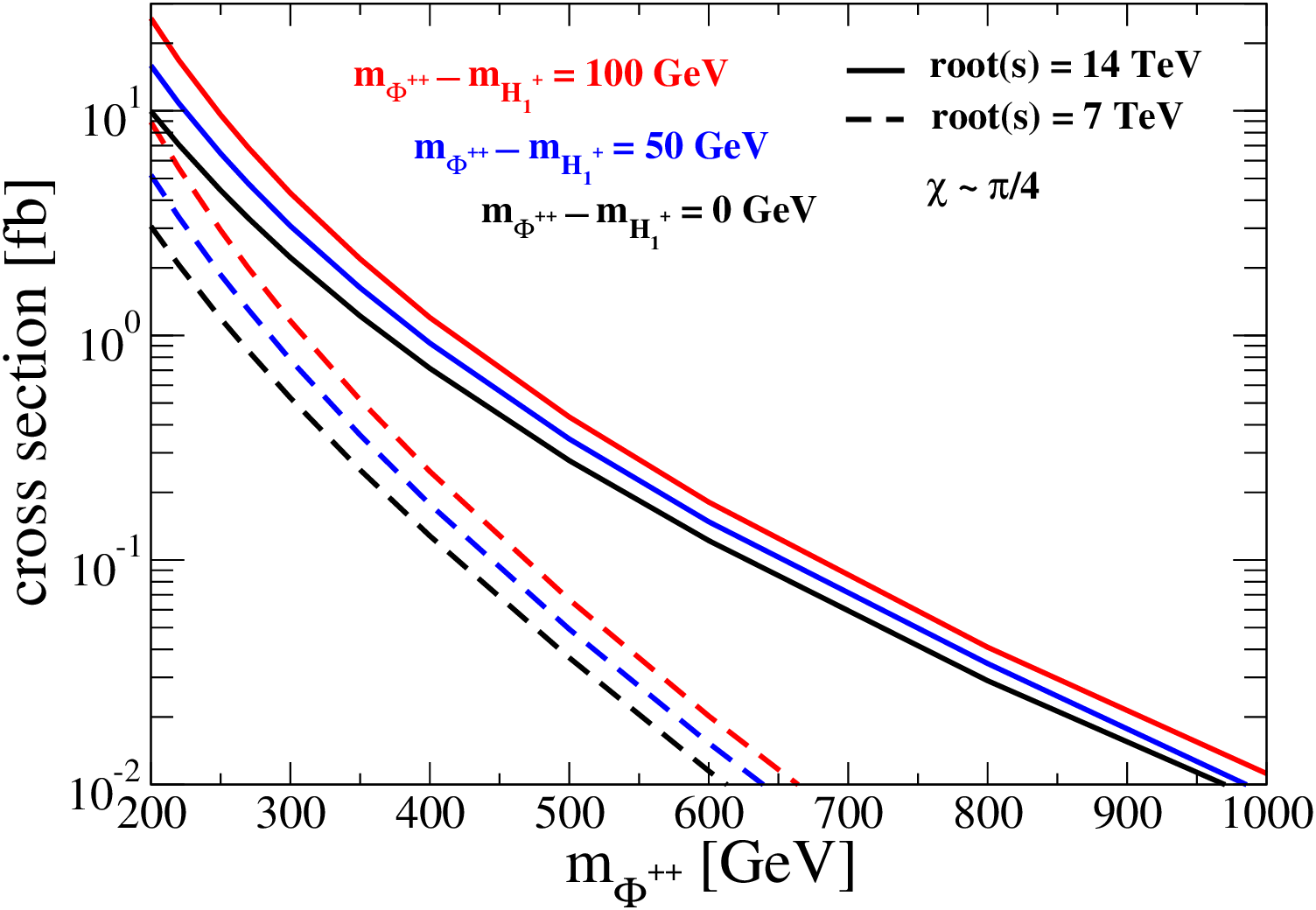}
\caption{Cross section of 
$pp \to X W^{+\ast} \to X \Phi^{++} H_1^-$. 
}\label{cs}
\end{center}
\end{figure}

The most general scalar potential is given by 
\begin{align}
&V=\sum_{i=1}^2 \mu_i^2|\phi_i|^2+(\mu_3^2\phi_1^\dagger \phi_2+\text{h.c.})
+\sum_{i=1}^2\frac{1}{2}\lambda_i|\phi_i|^4 \notag\\
&+\lambda_3|\phi_1|^2|\phi_2|^2+\lambda_4|\phi_1^\dagger\phi_2|^2
+\frac{1}{2}[\lambda_5(\phi_1^\dagger\phi_2)^2+\text{h.c.}]\notag\\
&+\mu_\Phi^2 |\Phi_{3/2}|^2+\frac{1}{2}\lambda_\Phi|\Phi_{3/2}|^4
+\sum_{i=1}^2 \rho_i|\phi_i|^2|\Phi_{3/2}|^2 \notag\\
&+\sum_{i=1}^2 \sigma_i |\phi_i^\dagger\Phi_{3/2}|^2 
+[\kappa(\Phi_{3/2}^\dagger\phi_1)(\phi_2\cdot \phi_1)+\text{h.c.}],
\label{potential}
\end{align}
where the $Z_2$ symmetry is softly broken at the $\mu_3^2$ term.  
We neglect the CP violating phase for simplicity.
The scalar doublets $\phi_1$, $\phi_2$ and $\Phi_{3/2}$ are 
parameterized as 
\begin{align}
\hspace{-0.2cm}\phi_i\!=\! \left[\begin{array}{c}
w_i^+ \\
\frac{1}{\sqrt{2}}(h_i+v_i+iz_i)
\end{array}\right]  (i=1,2), \; 
\Phi_{3/2} \!=\! \left[\begin{array}{c}
\Phi^{++} \\
\Phi^+
\end{array}\right], \nonumber
\end{align}
where the VEVs $v_i$ satisfy $\sqrt{v_1^2+v_2^2} = v \simeq 246$ GeV. 
Mass matrices for the neutral components are diagonalized 
as in the same way as 
those in the usual two Higgs doublet model (2HDM) with $\phi_1$ and $\phi_2$.
The mass eigenstates $h$ and $H$ for CP-even states are 
obtained by diagonalizing the mass matrix by the angle $\alpha$.  
By the angle $\beta$ ($\tan\beta \equiv v_2/v_1$), 
the mass eigenstates for the CP-odd states $z$ and $A$ 
are obtained, where $z$ is the Nambu-Goldstone (NG) boson 
and $A$ is the CP-odd Higgs boson. 
For simplicity $\sin(\beta-\alpha)=1$ is taken such that 
$h$ is the SM-like Higgs boson\cite{dec-2hdm}.
The existence of $\Phi_{3/2}$ affects the singly charged scalar sector. 
The mass eigenstates 
are obtained by mixing angles $\beta$ and $\chi$ as 
\begin{align}
\left[
\begin{array}{c}
w^\pm\\
H_1^\pm\\
H_2^\pm
\end{array}\right] 
=  
\left[
\begin{array}{ccc}
1 & 0 &0\\
0 & c_\chi & s_\chi \\
0&  -s_\chi &  c_\chi
\end{array}
\right] 
\left[
\begin{array}{ccc}
c_\beta & s_\beta & 0\\
-s_\beta & c_\beta & 0 \\
0&  0          & 1
\end{array}\right]
\left[
\begin{array}{c}
w_1^\pm\\
w_2^\pm\\
\Phi^\pm
\end{array}\right], \label{m+}
\end{align}
where $c_\theta=\cos\theta$ and $s_\theta=\sin\theta$, 
$w^\pm$ are the NG bosons   
absorbed by the longitudinal component of the $W^\pm$ bosons. 
$H_1^\pm$ and $H_2^\pm$  are physical mass eigenstates with 
the masses $m_{H_1^\pm}$ and $m_{H_2^\pm}$.

The Yukawa couplings for charged states are given by  
\begin{align}
&\mathcal{L}_Y = 
-\frac{\sqrt{2} V^{ij}_{\rm KM}}{v}
\bar{u}^i (m_{u^i} \xi_A^u d_L^j +m_{d^j}\xi_A^d d_R^j) \phi^+ 
\notag \\&
-\frac{\sqrt{2} m_{\ell^i}\xi_A^\ell}{v}\bar{\nu}^i \ell_R^i
\phi^+
  +\text{h.c.}, \;\; (\phi^+=H_1^+\cos\chi-H_2^+\sin\chi), \nonumber
\end{align}
where the coupling parameters $\xi^{u,d,\ell}_A$~\cite{typeX} 
depend on the $Z_2$ charges of quarks and leptons~\cite{grossman}.
We are interested in the light charged scalar bosons 
such as $\mathcal{O}(100)$ GeV. 
To satisfy the $b\to s\gamma$ data~\cite{bsg}, 
we choose 
the Type-I Yukawa interaction with $\tan\beta\gtrsim 2$. 
Assuming $m_{H_{1,2}^\pm} < m_t+m_b$ and $m_{H^\pm_2}-m_{H^\pm_1} < m_Z$,  
the branching ratios for the main decay modes are evaluated as 
$B(H_{1,2}^\pm \to \tau^\pm \nu) \sim 0.7$ 
and $B(H_{1,2}^\pm \to cs) \sim 0.3$ when $\chi \neq  0$.  

\begin{figure}[t]
\begin{center}
\includegraphics[width=42mm]{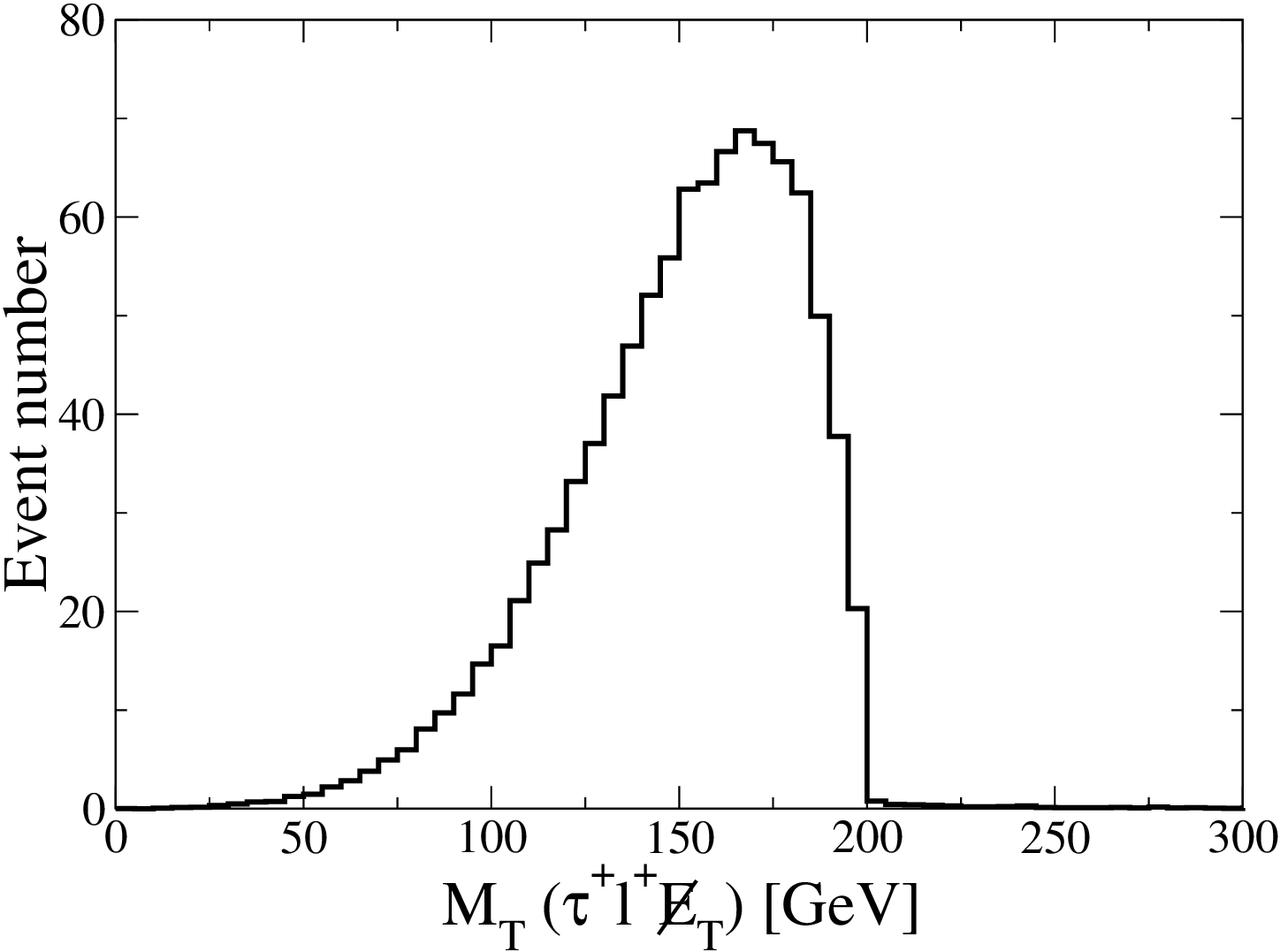} %\hspace{2mm}
\includegraphics[width=42mm]{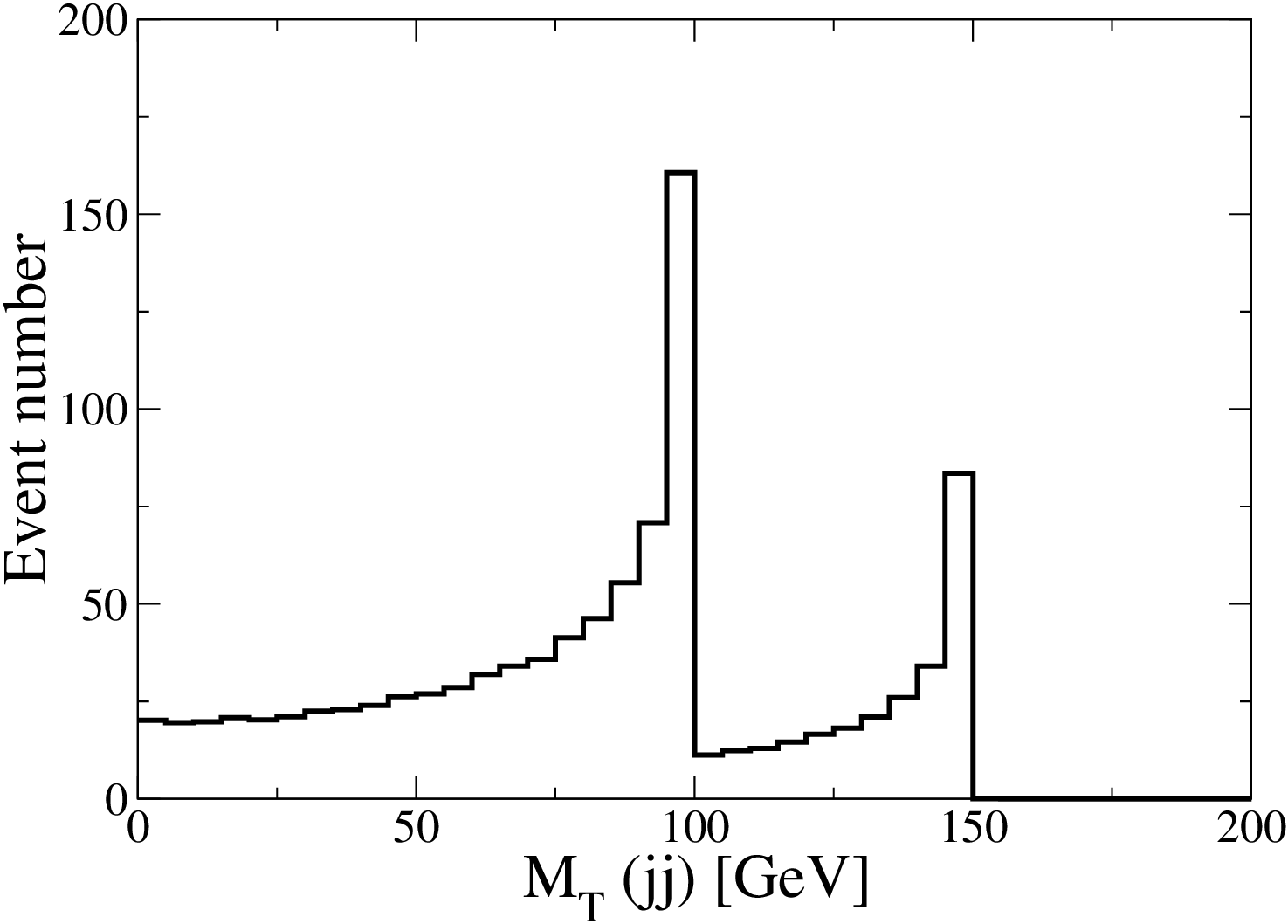}
\caption{(Left) 
The transverse mass distribution for the 
$\tau^+\ell^+E_T {\!\!\!\!\!\!\!/} \hspace{3mm}$ system 
for the signal. 
(Right) That for the $jj$ system. 
The event number is taken to be 1000 for both figures. 
}\label{fig_jacobian-peak}
\end{center}
\end{figure}

At the LHC, $\Phi^{++}$ can be tested by using 
various processes such as the pair production and  
the associated production with $H_1^{-}$ or $H_2^{-}$. 
We here discuss an interesting signal 
via the process $u \overline{d} \to W^{+\ast} \to \Phi^{++} H_{1,2}^-$. 
The cross section is shown in Fig.~\ref{cs}.   
We may examine this process, for example, by the decay 
$\Phi^{++} \to H_{1,2}^+W^+ \to \tau^+\ell^+\nu\nu$ with
$H_{1,2}^- \to jj$, 
when $m_{\Phi^{\pm\pm}} > m_{H_1^{\pm}} + m_W$ with 
$m_{H_{1,2}^\pm} < m_t+m_b$ and $m_{H_2^{\pm}} - m_{H_1^{\pm}} < m_Z$, 
where $m_{\Phi^{\pm\pm}}$ is the mass of $\Phi^{\pm\pm}$.     
The signal is then $\tau^+\ell^+ jj$ plus a missing transverse 
momentum $E_T {\!\!\!\!\!\!\!/}\hspace{3mm}$ ($\ell^+=e^+$ or $\mu^+$). 
The signal cross section for 
$\tau^\pm\ell^\pm jj E_T {\!\!\!\!\!\!\!/}\hspace{3mm}$ 
is evaluated as 
4.0 fb (1.3 fb) for $\sqrt{s} 
=14$ TeV ($7$ TeV) 
for $m_{H_1^\pm}=100$ GeV, $m_{H_2^\pm}=150$ GeV, 
$m_{\Phi^{\pm\pm}}=200$ GeV and $\chi \simeq \pi/4$.

The mass for $\Phi^{++}$ can be determined from 
the Jacobian peak~\cite{Jacobian} in the distribution of the 
transverse mass, 
%\begin{align}
$M_T(\tau^+\ell^+E_T {\!\!\!\!\!\!\!/} \hspace{3mm}) 
= \sqrt{2 p_T^{\tau\ell} E_T {\!\!\!\!\!\!\!/} \hspace{3mm}
(1-\cos\varphi)}$, 
%\end{align}
where $\varphi$ is the azimuthal angle between the transverse  
momentum $p_T^{\tau\ell}$ of the dilepton system and 
$E_T {\!\!\!\!\!\!\!/}\hspace{3mm}$. 
We show numerical results for the scenario with 
$m_{H_1^\pm}=100$ GeV, $m_{H_2^\pm}=150$ GeV, $m_{\Phi^{\pm\pm}}=200$ GeV, 
$\chi \simeq \pi/4$, $\tan\beta=3$, $\sin(\beta-\alpha)=1$ and 
$m_{H} = m_A = 127$ GeV, where $m_{H}$ and $m_A$ represent 
the masses of $H$ and $A$, respectively. 
The potential is then approximately custodial symmetric, 
so that the rho parameter constraint is satisfied with 
the mass of the SM-like Higgs boson $h$ to be 120 GeV. 
The end point in Fig.~\ref{fig_jacobian-peak}~(Left) 
indicates $m_{\Phi^{\pm\pm}}$, where the event number is 
taken to be 1000.   
One might think that the final decay products 
from the $\tau$ lepton should be discussed.  
We stress that the endpoint at $m_{\Phi^{++}}$ 
also appears in the distribution of 
$M_T(\ell^+\ell^+E_T {\!\!\!\!\!\!\!/} \hspace{3mm})$  
obtained from the leptonic decay of the $\tau^+$. 
The cross section for the 
$\ell^+\ell^+jjE_T {\!\!\!\!\!\!\!/} \hspace{3mm}$ signal 
is about 1.3 fb for $\sqrt{s}=14$ TeV (0.45 fb for $\sqrt{s}=7$ TeV).  
Furthermore, masses of singly charged Higgs bosons can also be 
measured by the distribution of $M_T(jj)$. %for the $jj$ system.  
In Fig.~\ref{fig_jacobian-peak}~(Right), 
the two Jacobian peaks at 100 and 150 GeV 
correspond to $m_{H_1^+}$ and $m_{H_2^+}$, respectively, where 
the event number is taken to be 1000. 
The SM background for $\ell^+\ell^+jjE_T {\!\!\!\!\!\!\!/} \hspace{3mm}$, 
which mainly comes from $u\bar{d}\to W^+W^+ jj$, 
is 3.95 fb for $\sqrt{s}= 14$ 
TeV (0.99 fb for $\sqrt{s}=7$ TeV). The cross section of the  
background is comparable to that for the signal before kinematic cuts. 
There is no specific kinematical structure in the $\ell^+\ell^+$ 
distribution in the background.   
All the charged scalar states can be measured simultaneously 
via this process unless their masses are too heavy 
if sufficient number of the signal event 
remains after kinematic cuts. 
While the detection at the LHC with 300 fb$^{-1}$ may be challenging, 
it could be much better at the upgraded version of the LHC with 3000 fb$^{-1}$.  

\begin{figure}[t]
\begin{center}
\includegraphics[width=60mm]{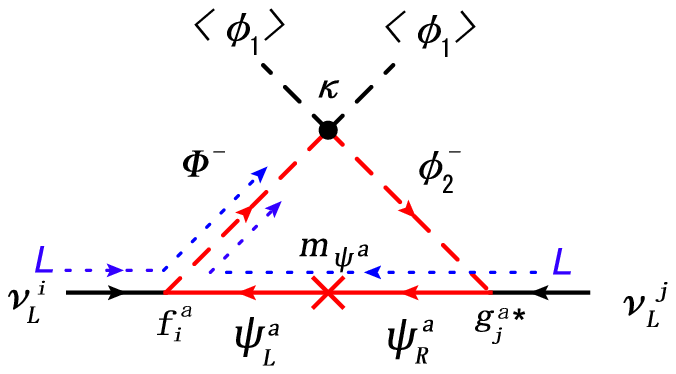}
\caption{Neutrino mass diagram.}\label{n_mass}
\end{center}
\end{figure}

\section{Model II} 
%\underline{\bf Model II.}  
We here present a new model   
in which $\Phi_{3/2}$ is introduced to naturally generate 
tiny neutrino masses at one-loop level. 
To this end, we again consider the scalar sector 
with $\phi_1$, $\phi_2$ and $\Phi_{3/2}$. 
In addition, we introduce two isospin singlet Dirac fermions $\psi^a$ 
($a=1,2$) with $Y=-1$. 
We impose the exact (unbroken) $Z_2$ parity, under which 
$\phi_2$, $\Phi_{3/2}$ and $\psi^a$ are odd while 
all the SM particles including $\phi_1$ are even.  
This $Z_2$ parity plays a role to forbid 
mixing terms of $\overline{\ell}_R \psi_L^a$ 
as well as couplings of $\overline{L}_L \phi_1 \psi_R^a$ and 
$\overline{L}_L \phi_2 \ell_R$, and to 
 guarantee the stability of a dark matter candidate; 
i.e., the lightest neutral $Z_2$ odd particle. 
Lepton numbers $L=-2$ and $+1$ are respectively assigned to 
$\Phi_{3/2}$ and $\psi^a$.  

The scalar potential coincides that in Eq.~(\ref{potential})  
but $\mu_3^2=0$ due to the exact $Z_2$ parity. 
Without $\Phi_{3/2}$, 
the scalar sector is that of the  
inert doublet model~\cite{inert}, in which only $\phi_1$ receives 
the VEV yielding the SM-like Higgs boson $h$,  
while $\phi_2$ gives $Z_2$-odd scalar bosons $H$, $A$ and $H^\pm$. 
Including $\Phi_{3/2}$, 
$H^\pm$ can mix with $\Phi^{\pm}$ 
diagonalized by the angle $\chi$ 
in Eq.~(\ref{m+}) with $\beta=0$.
Masses and interactions for $\psi^a$ are given by 
\begin{align}
\mathcal{L}_Y= %\sum_{a=1}^2
m_{\psi^a} \bar{\psi}_L^a \psi_R^a
+f_i^a\overline{(L_L^i)^c}\cdot\Phi_{3/2}\psi_L^a
+g_i^a\overline{L_L^i}\phi_2\psi_R^a + {\rm h.c.}. \label{yukawa_model2}
\end{align}

The neutrino masses are generated via the one-loop diagram 
in Fig.~\ref{n_mass}. 
The flow of the lepton number is also indicated in the figure. 
The source of lepton number violation (LNV) is the coupling $\kappa$. 
This is similar to the model by Zee~\cite{zee}, although 
the diagram looks similar to the model by Ma~\cite{ma} where 
Majorana masses of right-handed neutrinos $\nu_R$ 
is the origin of LNV.   
For $m_{\psi^a} \gg m_{H_1^\pm},m_{H_2^\pm}$, the mass matrix can be calculated as 
\begin{align}
(\mathcal{M}_\nu)_{ij}&\simeq \sum_{a=1}^2
\frac{1}{16\pi^2}\frac{1}{2m_{\psi^a}}(f_i^a g_j^{a*}+f_j^a g_i^{a*})
\frac{v^2\kappa}{m_{H_2^\pm}^2-m_{H_1^\pm}^2}\notag\\
& \times \left(m_{H_2^\pm}^2\log\frac{m_{\psi^a}^2}{m_{H_2^\pm}^2}
-m_{H_1^\pm}^2\log\frac{m_{\psi^a}^2}{m_{H_1^\pm}^2}\right). \nonumber
\end{align} 
For $m_\psi\sim 1$ TeV, 
$m_{H_1^\pm} \sim m_{H_2^\pm} \sim {\cal O}(100)$~GeV, and 
$f_i^a \sim g_i^a \sim \kappa \sim {\cal O}(10^{-3})$, 
the scale of neutrino masses ($\sim 0.1$ eV) can 
be generated.  The bound from LFV processes  
such as $\mu\to e\gamma$~\cite{meg} can easily be satisfied.
The neutrino data can be reproduced by introducing at least 
two fermions $\psi^1$ and $\psi^2$.
The lightest $Z_2$ odd neutral Higgs boson
(either $H$ or $A$) is a dark matter candidate~\cite{inert}. 
Assuming that $H$ is the lightest,  
its thermal relic abundance
can explain the WMAP data~\cite{wmap} 
by the s-channel process 
$HH\to h \to b \overline{b}$ (or $\tau^+\tau^-$).  
The t-channel process 
$HH\to \overline{\ell}_L \ell_L$ with $\psi_R$ 
mediation 
is negligible. 
The direct search results can also be satisfied. 

Finally, we comment on the collider signature in Model~II. 
$\Phi_{3/2}$ is $Z_2$ odd,  
so that its decay product includes the dark matter $H$. 
For $m_H=50$ GeV, the mass of $h$  
would be about 115~GeV to satisfy the WMAP data~\cite{wmap}.  
We then consider the parameter set; 
$m_{\Phi^{++}}=230$ GeV, $m_{H_2^+}=150$ GeV,  
$m_{H_1^+} \simeq m_A = 149$ GeV and $\chi=0.1$  
to satisfy the neutrino data and the LFV data. 
The signal at the LHC would be 
$W^+W^+W^-E_T {\!\!\!\!\!\!\!/} \hspace{3mm}$ via 
$u \overline{d} \to \Phi^{++} H_{i}^- 
\to (H_{i}^+ W^+)(W^- H) \to (H W^+W^+)(W^- H)$. 
The cross section of 
$W^\pm W^\pm W^\mp E_T {\!\!\!\!\!\!\!/} \hspace{3mm}$
is 23 fb for $\sqrt{s}=14$ TeV (7.3 fb for $\sqrt{s}=7$ TeV).
The main background comes from $W^\pm W^\pm W^\mp$, 
and the cross section 
is 135 fb for $\sqrt{s}=14$ TeV (76 fb for $\sqrt{s}=7$ TeV). 
The signal background ratio is not too small at all, 
and we can expect the signal would be detected after 
appropriate kinematic cuts.

\section{Discussions and conclusions } 
%\underline{\bf Model~III}
Let us discuss the exact (unbroken) $Z_2$ parity 
%imposed 
in Model II. 
In radiative seesaw models with $\nu_R$~\cite{knt,ma,aks},  
the Majorana masses are the source of LNV, and 
an exact $Z_2$ has to be imposed to forbid 
the neutrino Yukawa coupling. 
On the contrary, in Model II, the source 
of LNV is the $\kappa$ term in Eq.~(\ref{potential}), 
so that  
the canonical seesaw mechanism~\cite{see-saw} 
cannot occur even without such an exact $Z_2$ parity.
Thus, in Model II the exact $Z_2$ parity is not necessarily 
important for radiative generation of neutrino masses.  
Therefore, we may consider another model (Model III), 
in which the exact $Z_2$ parity is softly broken in Model~II.
Then, in addition to the terms in Eq.~(\ref{yukawa_model2}),  
new terms of $\overline{\psi}_L^a \ell_R^i$ 
appear.  
They cause LFV processes such as $\mu \to eee$ or $\mu\to e\gamma$.  
By setting $m_{\psi^a}$ to be at TeV scales 
with smaller mixing parameters 
$\tilde{m}_{\psi\ell}^{a i}$ of $\overline{\psi}^a_L \ell_R^i$,  
such LFV processes can be suppressed 
to satisfy the current data. 
Note that the tree level LFV process such as 
$\mu\to eee$ via $\overline{\psi}_L$-$\ell_R$ mixing is multiplied by
a suppression factor like $m_\ell^2\tilde{m}_{\psi\ell}^2/(v^2 m_\psi^2)$. 
%, 
Neutrino masses are generated by not only 
the diagram in Fig.~\ref{n_mass} but also 
additional diagrams where $\ell_R^i$ instead of $\psi_R^a$ 
and $\phi^-$ instead of $\phi_2^-$ are in the loop.
Phenomenology of the Higgs sector in Model~III  coincides 
that in Model~I, so that the same signal shown in  
Fig.~\ref{fig_jacobian-peak} can be used to 
identify the $\Phi_{3/2}$ field.  

We have briefly discussed 
collider phenomenology of $\Phi_{3/2}$ 
in Models~I to III.
We comment on the 
difference from models with $S^{++}$  or 
$\Delta=(\Delta^{++}, \Delta^{+}, \Delta^{0})$.  
First, the production process 
$u \overline{d} \to W^{+\ast} \to 
\varphi^{++} \varphi^-$ ($\varphi$: a scalar component)
can be useful to test $\Phi_{3/2}$ and $\Delta$~\cite{aa}, 
while $S^{++}$ does not contribute 
to this process because of no weak gauge coupling. 
In the model with a $\Delta$, although the signal strongly 
depends on the parameters~\cite{tao},  
$\ell^+\ell^+ jj E_T {\!\!\!\!\!\!\!/}\hspace{3mm}$ 
can be important for $m_{\Delta^{++}}-m_{{\Delta}^+} > {\cal O}(1)$~GeV 
when the hadronic decay mode of the singly charged state 
is substantial. 
In this case, masses of doubly and singly charged scalars 
can be measured from the end points in 
$M_T(\ell^+\ell^+ E_T {\!\!\!\!\!\!\!/}\hspace{3mm})$ 
and $M_T(jj)$ distributions, similarly to Models I and III.
Otherwise, a peak in the invariant mass distribution  
$M(\ell^+\ell^+)$ would be seen at 
$m_{\Delta^{++}}$ in the triplet models when the VEV of 
the triplet field is sufficiently small~\cite{tao}, 
while there is no such a peak in the model with $\Phi_{3/2}$. 
Second, the production mechanism $u\overline{d} \to W^{+\ast} \to 
\varphi^+\varphi^0$ can be useful to study multi-doublet 
models~\cite{ky} and the model with $\Delta$ as well as 
that with $\Phi_{3/2}$ %if there is mixing 
with an extra doublet.  
There is no such process in the model with $S^\pm$.  
The signal strongly depends on the model; i.e., 
$bb\tau E_T {\!\!\!\!\!\!\!/}\hspace{3mm}$ 
for the Type-II 2HDM~\cite{ky}, 
$\tau\tau\tau E_T {\!\!\!\!\!\!\!/}\hspace{3mm}$ for the 
Type-X 2HDM~\cite{typeX}, 
and some others in the model with $\Delta$. 
Therefore, extracting properties of doubly and singly 
charged Higgs bosons by using 
these processes as well as  
other various specific processes, 
we can discriminate the Higgs sectors at the LHC 
(and its luminosity-upgraded version) 
to a considerable extent. 

%
%\section{Conclusion} 

To summarize, we have studied various aspects of $\Phi_{3/2}$ including  
the signature at the LHC in a few models.     
New TeV-scale models with $\Phi_{3/2}$ have been 
presented for generating tiny neutrino masses, one of which 
also contains dark matter candidates. 
We have found that $\Phi_{3/2}$ in these models 
shows discriminative and 
testable aspects at the LHC and its luminosity-upgraded version, 
so that models with $\Phi_{3/2}$ 
would be distinguishable from the other models 
with doubly charged scalar states.
The details are discussed elsewhere~\cite{aky2}.

%

%{\bf Acknowledgments}~~~\\[2mm]
%\indent
This work was supported by Grant-in-Aid for Young Scientists (B) 
no. 22740137 (M.A.), and Grant-in-Aid for Scientific Research (A) no. 
22244031 (S.K.). 
K.Y. was supported by Japan Society for the Promotion of Science. 
%(JSPS Fellow (DC2)).

Note Added: after this Letter was completed, 
a paper~\cite{su} appeared, where same sign dilepton 
resonances were discussed for $\Phi^{++}$ together 
with $\Delta^{++}$ and $S^{++}$.


\vspace*{-4mm}
\begin{thebibliography}{1}

\bibitem{LRmodel} 

  R.~N.~Mohapatra and G.~Senjanovic,
  %``Neutrino Mass and Spontaneous Parity Violation,''
  Phys.\ Rev.\ Lett.\  {\bf 44}, 912 (1980).
  %%CITATION = PRLTA,44,912;%%

\bibitem{LHmodel}

T.~Han, H.~E.~Logan, B.~McElrath and L.~T.~Wang, 
  %``Phenomenology of the little Higgs model,''
  Phys.\ Rev.\  D {\bf 67}, 095004 (2003).
%  [arXiv:hep-ph/0301040]. 

\bibitem{TypeIIseesaw}

  J.~Schechter and J.~W.~F.~Valle,
  %``Neutrino Masses in SU(2) x U(1) Theories,''
  Phys.\ Rev.\  D {\bf 22}, 2227 (1980); 
  G.~B.~Gelmini and M.~Roncadelli,
  %``Left-Handed Neutrino Mass Scale and Spontaneously Broken Lepton Number,''
  Phys.\ Lett.\  B {\bf 99}, 411 (1981).

\bibitem{ZBmodel} 

 A.~Zee,
 %``Quantum Numbers Of Majorana Neutrino Masses,''
 Nucl.\ Phys.\ B {\bf 264}, 99 (1986);
 K.~S.~Babu,
 %``MODEL OF 'CALCULABLE' MAJORANA NEUTRINO MASSES,''
 Phys.\ Lett.\ B {\bf 203}, 132 (1988).

\bibitem{gunion} 
J.~F.~Gunion, hep-ph/0212150.

\bibitem{knt}
  L.~M.~Krauss, S.~Nasri and M.~Trodden,
  %``A model for neutrino masses and dark matter,''
  Phys.\ Rev.\  D {\bf 67}, 085002 (2003); 

\bibitem{ma}
  E.~Ma,
  %``Verifiable radiative seesaw mechanism of neutrino mass and dark matter,''
  Phys.\ Rev.\  D {\bf 73}, 077301 (2006). 

\bibitem{aks}
  M.~Aoki, S.~Kanemura and O.~Seto,
  %``Neutrino mass, Dark Matter and Baryon Asymmetry via TeV-Scale Physics
  %without Fine-Tuning,''
  Phys.\ Rev.\ Lett.\  {\bf 102}, 051805 (2009).

\bibitem{zee}
  A.~Zee,
  %``A Theory Of Lepton Number Violation, Neutrino Majorana Mass, And
  %Oscillation,''
  Phys.\ Lett.\  B {\bf 93}, 389 (1980)
  [Erratum-ibid.\  B {\bf 95}, 461 (1980)];
  A.~Zee,
  %``Charged Scalar Field And Quantum Number Violations,''
  Phys.\ Lett.\  B {\bf 161}, 141 (1985).

\bibitem{gunion2} 

  J.~F.~Gunion, C.~Loomis and K.~T.~Pitts,
  arXiv:hep-ph/9610237.

\bibitem{gw}
  S.~L.~Glashow and S.~Weinberg,
  %``Natural Conservation Laws For Neutral Currents,''
  Phys.\ Rev.\  D {\bf 15}, 1958 (1977).

\bibitem{dec-2hdm} 

  J.~F.~Gunion and H.~E.~Haber,
  %``The CP conserving two Higgs doublet model: The Approach to the decoupling
  %limit,''
  Phys.\ Rev.\  D {\bf 67}, 075019 (2003); 
  S.~Kanemura, Y.~Okada, E.~Senaha and C.~P.~Yuan,
  %``Higgs coupling constants as a probe of new physics,''
  Phys.\ Rev.\  D {\bf 70}, 115002 (2004).

\bibitem{typeX} 
  M.~Aoki, S.~Kanemura, K.~Tsumura and K.~Yagyu,
  %``Models of Yukawa interaction in the two Higgs doublet model, and their
  %collider phenomenology,''
  Phys.\ Rev.\  D {\bf 80} (2009) 015017.

\bibitem{grossman}

  V.~D.~Barger, J.~L.~Hewett and R.~J.~N.~Phillips,
  %``NEW CONSTRAINTS ON THE CHARGED HIGGS SECTOR IN TWO HIGGS DOUBLET MODELS,''
  Phys.\ Rev.\  D {\bf 41}, 3421 (1990);
  Y.~Grossman,
  %``Phenomenology of models with more than two Higgs doublets,''
  Nucl.\ Phys.\  B {\bf 426}, 355 (1994).

\bibitem{bsg} 

  D.~Asner {\it et al.}, % [Heavy Flavor Averaging Group],
  %``Averages of b-hadron, c-hadron, and $\tau-lepton Properties,''
  arXiv:1010.1589 [hep-ex].
  %%CITATION = ARXIV:1010.1589;%%
\bibitem{Jacobian} 
  J. Smith, W.~L.~van~Neerven, and J.~A.~M.~Vermaseren, 
  Phys.\ Rev.\ Lett.\  {\bf 50}, 1738 (1983).

\bibitem{inert} 
  L.~Lopez Honorez, E.~Nezri, J.~F.~Oliver and M.~H.~G.~Tytgat,
%  L.~Lopez Honorez, et al., 
  %``The inert doublet model: An archetype for dark matter,''
  JCAP {\bf 0702} (2007) 028; 
    Q.~H.~Cao, E.~Ma and G.~Rajasekaran,
  %``Observing the Dark Scalar Doublet and its Impact on the Standard-Model
  %Higgs Boson at Colliders,''
  Phys.\ Rev.\  D {\bf 76} (2007) 095011; 
%\cite{Dolle:2009fn}
%\bibitem{Dolle:2009fn}
  E.~M.~Dolle and S.~Su,
  %``The Inert Dark Matter,''
  Phys.\ Rev.\  D {\bf 80}, 055012 (2009).
%  [arXiv:0906.1609 [hep-ph]].
  %%CITATION = PHRVA,D80,055012;%%


\bibitem{meg} 
%\cite{Adam:2009ci}
%\bibitem{Adam:2009ci}
  J.~Adam {\it et al.},  %[MEG collaboration],
  %``A limit for the mu ---> e gamma decay from the MEG experiment,''
  Nucl.\ Phys.\  B {\bf 834}, 1 (2010). 
%  [arXiv:0908.2594 [hep-ex]].
  %%CITATION = NUPHA,B834,1;%%


\bibitem{wmap} 
%\cite{Komatsu:2010fb}
%\bibitem{Komatsu:2010fb}
  E.~Komatsu {\it et al.},   %[WMAP Collaboration],
  %``Seven-Year Wilkinson Microwave Anisotropy Probe (WMAP) Observations:
  %Cosmological Interpretation,''
  Astrophys.\ J.\ Suppl.\  {\bf 192}, 18 (2011). 
%  [arXiv:1001.4538 [astro-ph.CO]].
  %%CITATION = APJSA,192,18;%%


\bibitem{see-saw}
         T.~Yanagida, In Proceedings of Workshop on  {\it the Unified
        Theory and the Baryon Number in the Universe}, p.95 KEK Tsukuba,
        Japan (1979);
        M. Gell-Mann, P.~Ramond and R.~Slansky, 
        in Proceedings of Workshop {\it Supergravity}, p.315, Stony
        Brook, New York, 1979.

\bibitem{aa}
  A.~G.~Akeroyd and M.~Aoki,
  %``Single and pair production of doubly charged Higgs bosons at hadron
  %colliders,''
  Phys.\ Rev.\  D {\bf 72}, 035011 (2005).
  %[arXiv:hep-ph/0506176].

\bibitem{tao}
  P.~Fileviez~P\'{e}rez, et al, Phys.\ Rev.\ D {\bf 78} 015018 (2008).

\bibitem{ky}
  S.~Kanemura and C.~P.~Yuan,
  %``Testing supersymmetry in the associated production of CP-odd and  charged
  %Higgs bosons,''
  Phys.\ Lett.\  B {\bf 530}, 188 (2002);
  Q.~H.~Cao, S.~Kanemura and C.~P.~Yuan,
  %``Associated production of CP-odd and charged Higgs bosons at hadron
  %colliders,''
  Phys.\ Rev.\  D {\bf 69}, 075008 (2004); 
%  A.~Belyaev, Q.~-H.~Cao, D.~Nomura, K.~Tobe, C.~-P.~Yuan,
  A.~Belyaev, et al., 
  %``Light MSSM Higgs boson scenario and its test at hadron colliders,''
  Phys.\ Rev.\ Lett.\  {\bf 100}, 061801 (2008).

\bibitem{aky2} 
   M.~Aoki, S.~Kanemura, and K.~Yagyu, in preparation.

\bibitem{su} 
  V.~Rentala, W.~Shepherd, and S.~Su, arXiv:1105.1379.

\end{thebibliography}
\end{document}